\documentclass[12pt]{JHEP3}

\usepackage{mathrsfs}
\usepackage{amsmath,amssymb}
\usepackage{epsfig}
\input epsf

\def\l{{\lambda}}
 \def \const {{\rm const}}

\def\be{\begin{equation}}
\def\ee{\end{equation}}

\newcommand{\bea}{\begin{eqnarray}}
\newcommand{\eea}{\end{eqnarray}}

\def\s {\sigma}
\def\pa {\partial}

\def\la{\label}

\def\ov{\over}

\newcommand{\alg}[1]{\mathfrak{#1}}
\newcommand{\su}{\alg{su}}

\newcommand{\psu}{\alg{psu}}

\newcommand{\AdS}{{\rm  AdS}_5\times {\rm S}^5}

\newcommand{\sfrac}[2]{{\textstyle\frac{#1}{#2}}}

\def\ba{\begin{array}}
\def\ea{\end{array}}

\def \const {{\rm const}}

\def\s{\sigma}

\def\ov{\over}

\def\l{\lambda}
\def\eps{\zeta}

\def \l {\lambda}

\def \ov {\over}

\def \ss {\alg{s}}

\preprint{  {\tt hep-th/0604043}\\ {\tt AEI-2006-022}
\\ {\tt  ITP-UU-06-15}
\\{\tt SPIN-06-13}
}

\title{On
 $\AdS$ String S-matrix}

\author{
G. Arutyunov$^{a}$\footnote{e-mails: G.Arutyunov@phys.uu.nl, frolovs@aei.mpg.de}
\footnote{Correspondent fellow at Steklov Mathematical Institute,
Moscow } and S. Frolov$^{b}$\footnote{Correspondent fellow at Steklov Mathematical Institute,
Moscow }
\\
\\
$^{a}$ {\it Institute for Theoretical Physics and Spinoza
Institute,
Utrecht University \\
~~3508 TD Utrecht, The Netherlands}\\
$^{b}$ {\it Max-Planck-Institut f\"ur Gravitationsphysik,
Albert-Einstein-Institut\\
~~Am M\"uhlenberg 1,
D-14476 Golm, Germany }\\
 }

\abstract{Recently two interesting conjectures about the string
S-matrix on $\AdS$ have been made. First, assuming the existence
of a Hopf algebra symmetry Janik derived a functional equation for
the dressing factor of the quantum string Bethe ansatz. Second,
Hern\'andez and L\'opez proposed an explicit form of $1/\sqrt\l$
correction to the dressing factor. In this note we show that in
the strong coupling expansion Janik's equation is solved by the
dressing factor up to the order of its validity. This observation
provides a strong evidence in favor of a conjectured Hopf algebra
symmetry for strings in $\AdS$ as well as the perturbative string
S-matrix.
 }

\begin{document}
The S-matrix of the quantum string Bethe ansatz \cite{AFS}
coincides with the S-matrix of the asymptotic ${\cal N}=4$ SYM
Bethe ansatz \cite{MZ,BDS} up to a scalar function called the
dressing factor. It appears to be universal for all sectors
\cite{S,B}. The leading form of the dressing factor at large $\l$
was determined by discretizing the integral equations \cite{KMMZ}
which describe the spectrum of spinning strings in the scaling
limit of \cite{FT}. The analysis of one-loop corrections to
energies of spinning strings \cite{FT2} revealed  that the
dressing factor acquires $1/\sqrt\l$ corrections
\cite{SZZ}-\cite{SZ}. The explicit form of these corrections was
then conjectured in \cite{HL}.


Recently Janik put forward a proposal \cite{J} that a gauged-fixed
string sigma model on a plane exhibits a Hopf algebra symmetry
which allows one to derive a set of functional equations on the
dressing factor. The action of the Hopf algebra antipode is an
analog of the particle-to-antiparticle transformation in
relativistic field theory \cite{Cross}.  Implementing the antipode
action in a given representation of the Lie algebra
$\su(2|2)\oplus \su(2|2)$ leads to nontrivial relations for the
corresponding S-matrix. These relations are analogous to the
crossing symmetry relations which arise in relativistic integrable
models \cite{ZZ}.

The construction of \cite{J} implicitly assumes that the string
sigma model is quantized in a gauge preserving the invariance
under the $\su(2|2)\oplus \su(2|2)$ subalgebra of $\psu(2,2|4)$,
the latter being the symmetry algebra of the string sigma model on
$\AdS$. The two copies of $\su(2|2)$ subalgebra share the same
central element which is the string Hamiltonian in the gauge
chosen. For instance, one can consider the string sigma model in
the temporal gauge $t=\tau\,,\ p_\phi =J$, where $p_{\phi}$ is the
canonical momentum conjugate to an angle variable $\phi$ of ${\rm
S}^5$ \cite{AF}. Another example is given by the uniform
light-cone gauge $x^{+}=\tau$ and $p^+={\const}$ \cite{AFlc}. In
these type of gauges the string Lagrangian depends on two
parameters, e.g. in the temporal gauge, it depends on the string
tension $\sqrt{\lambda}$ and $J$. For finite $\l$ and $J$ the
gauged-fixed theory is a two-dimensional model on a cylinder and
by this reason the notion of the S-matrix is not defined. On the
other hand, at infinite $J$ with $\lambda$ finite the gauge-fixed
string sigma model is described by a two-dimensional field theory
on a plane because the $J$-dependence of the string Lagrangian can
be absorbed into rescaling of the world-sheet $\s$-coordinate
\cite{AAF}. The rescaled range of $\s$ is $-\pi J\le\s\le\pi J$,
and in the limit $J\to\infty$ one gets a model on a plane. The
S-matrix for the model can be determined by using the symmetry
algebra (and choosing properly its representation) up to a scalar
factor \cite{B}. The functional equations of \cite{J} might be
further used to fix the scalar factor. Different solutions of the
functional equations would correspond to different gauge choices
respecting the residual $\su(2|2)\oplus \su(2|2)$ symmetry.

The resulting S-matrix is the main building block to derive a set
of Bethe equations along the lines of Ref.\cite{B}.  However, the
following is to be mentioned. First, the Bethe equations arising
in this way are asymptotic and they hardly capture the exact
spectrum of strings at finite $J$. In fact, additional exponential
corrections of the form $e^{-J/\sqrt\l}$, which are  due to
finite-size effects, are expected \cite{AJK}. Second, it is
possible that at finite $J$ and $\l$ the Bethe equations should be
abandoned for direct diagonalization of a short-range Hubbard type
Hamiltonian \cite{Hubbard}. Third, it is presently unclear if and
how the string Bethe equations turn into the {\it gauge theory}
asymptotic Bethe equations \cite{BDS} in the weak coupling limit
$\l\to 0$, $J$ fixed. One of the possibilities here is that
starting at 4-loop order of weak coupling perturbation theory the
dressing factor could lead to violation of the BMN \cite{BMN}
scaling. In this respect we note  that the breakdown of the BMN
scaling was indeed observed in the plane-wave matrix models
\cite{KP}.

In this letter we analyze the dressing factor taking into account
the $1/\sqrt\l$ correction suggested in \cite{HL}, and show that
it satisfies the functional equation  in the large $\l$ limit up
to the second order of perturbation theory. This result can be
considered as a nontrivial test of the both proposals of \cite{HL}
and \cite{J}.

\bigskip

To formulate the string and gauge theory Bethe equations it is
convenient to use the variables $x^\pm$ introduced in \cite{Bx},
which satisfy the following equation
$$
x^+ + \frac{\lambda}{16\pi^2 x^+}-x^- - \frac{\lambda}{16\pi^2 x^-} = i\,.
$$
The momentum $p$ of a physical excitation is expressed via $x^\pm$
as $e^{ip}=\frac{x^+}{x^-}$.

To study the strong coupling expansion it is useful to rescale
$x^\pm$ as follows $$ x^\pm\to \frac{\sqrt{\lambda}}{4\pi}
x^\pm\,. $$ Then the rescaled $x^\pm$ satisfy the relation $$ x^+
+ \frac{1}{x^+}-x^- - \frac{1}{ x^-} =
i\frac{4\pi}{\sqrt{\lambda}} = 2i\eps\,, $$ where we introduced
the notation  $ \eps = \frac{2\pi}{\sqrt{\lambda}}\,. $ In fact
$1/\zeta$ is equal to the effective string tension. We choose the
following parametrization of $x^\pm$ in terms of a unconstrained
variable\footnote{This variable should not be confused with the
variable $x$ used in \cite{Bx}.} $x$ \bea\la{xpm} x^\pm(x) = x
\sqrt{1 -  { \eps^2\ov (x - \frac 1x)^2}} \pm i \eps { x\ov x
-\frac 1x}\,, \eea The momentum $p$ is related to $x$  through $$
\sin {p\ov 2} = { \eps \ov x -\frac{1}{x} }\, , $$ and the energy
of a physical excitation is $$
e(x)=\frac{x+\frac{1}{x}}{x-\frac{1}{x}}\, . $$ An interesting
feature of the above formula is that in this parametrization the
energy does not explicitly depend on the coupling constant $\eps$.
A dependence on the coupling will arise upon solving the Bethe
equations to be discussed below. Also, as we will see later on,
the obvious singularity of these formulae at $x=1$ is related to
the branch cut singularity of the perturbative string S-matrix. It
is not difficult to verify that the particle-to-antiparticle
transformation, $x^\pm\to 1/x^\pm$, is just the inversion $x\to
1/x$ $$ x^\pm(1/x) = 1/x^\pm(x)\, $$ and it transforms $e(x)$ to
$-e(x)$.

\medskip

To fix the conventions we write down the Bethe ansatz equations
for rank-one sectors \bea \label{Bethe} e^{ip_j L}=\prod_{k\neq
j}^M S(x_j,x_k)\,. \eea Here the string S-matrix is given by \bea
S(x_j,x_k) = \left( \frac{x^+_j-x_k^-}{x^-_j-x_k^+} \right)^{\ss}
\frac{1-\frac{1}{x_j^+x_k^-}}{1-\frac{1}{x_j^-x_k^+}}
\s(x_j,x_k)\,  \eea and $M$ is a number of excitations (Bethe
roots), $L=J+\sfrac{\ss+1}{2}M$, where $J$ is a
$\alg{u}(1)$-charge, and $\ss=1,0,-1$ for $\su(2)$, $\su(1|1)$ and
$\alg{sl}(2)$ sectors respectively. We point out that the S-matrix
describes the scattering of string states in the temporal gauge
$t=\tau$ and $p_{\phi}=J$ in the limit $J\to\infty$ with $\lambda$
kept fixed.

 Finally, the
function $\s(x_j,x_k)$ appearing in the string S-matrix is called
the dressing factor. Being universal to all sectors, it cannot be
fixed by $\psu(2,2|4)$ symmetry and therefore is supposed to
encode dynamical information about the model. The dressing factor
depends on the coupling constant $\lambda$ and, according to the
AdS/CFT correspondence, should be equal to one at $\lambda=0$ to
recover the perturbative gauge theory results. On the other hand,
at strong coupling the dressing factor can be determined by
studying the spectrum of string theory states in the near
plane-wave limit or, alternatively, the spectrum of semi-classical
spinning strings. This analysis leads to the following structure
of the dressing factor $\s(j,k)$ \bea \label{dr} \s(x_j,x_k) =
e^{i\theta(x_j,x_k)}\, , \eea where the dressing phase is a
bilinear form\footnote{This functional form of the dressing factor
was found by analyzing the most general long-range integrable
deformations of XXX spin chains \cite{BK}.} of local excitation
charges $q_r$ \bea \theta(x_j,x_k) &=& {1\ov
\eps}\sum_{r=2}^\infty \sum_{n=0}^\infty c_{r,r+1+2n}(\eps)\left(
q_r(x_j)q_{r+1+2n}(x_k) - q_r(x_k)q_{r+1+2n}(x_j) \right)\,.
\label{dph}\eea The local charges are defined as follows
\bea\la{charge} q_r(x_k) = {i\ov r - 1}\left( \left({1\ov
x^+_k}\right)^{r - 1} - \left({1\ov x^-_k}\right)^{r -
1}\right)\,, \eea and the functions $c_{r,s}$ can be expanded in
power series in $\eps$, where the first two terms of this
expansion are \bea c_{r,s}(\eps) = \delta_{r+1,s} - \eps {4\ov
\pi}{(r-1)(s-1)\ov (r+s-2)(s-r)}+ \cdots\,. \eea Here the leading
term was found \cite{AFS} by discretizing the integral equations
describing the finite-gap solutions of the classical string
sigma-model \cite{KMMZ}. The subleading term was recently proposed
in \cite{HL} by studying the one-loop sigma model corrections to
circular spinning strings. It is worth noting that the expansion
for the dressing phase is not strictly speaking an expansion in
$\eps$ because the charges $q_r$ have non-trivial dependence on
$\eps$. Therefore, the strong coupling expansion requires also
expanding the charges $q_r$.

\medskip

The functional equations of \cite{J} were written for the function
$S_0$ which is related to the dressing factor (\ref{dr}) as
follows\footnote{It is worth mentioning that $S_0$ is equal to the
S-matrix for the $\alg{sl}(2)$ sector ($s=-1$).} \bea
S_0(x_j,x_k)&=&\frac{x_j^--x_k^+}{x_j^+-x_k^-}
\frac{1-\frac{1}{x_j^+x_k^-}}{1-\frac{1}{x_j^-x_k^+}}
\s(x_j,x_k)\, .
 \eea
The functional equation to be satisfied by $S_0$ is \cite{J}
 \bea\label{S0}
S_0(x_j,x_k)S_0(1/x_j,x_k)=f(x_j,x_k)^{-2}\,,
 \eea
where the function $f(x_j,x_k)$ is \bea
f(x_j,x_k)=\frac{1-\frac{1}{x_j^+x_k^-} }{1-\frac{1}{x_j^-x_k^-}}
\frac{x_j^+-x_k^+}{x^-_j-x_k^+}\,.\eea It follows from
eq.(\ref{S0}) that $S_0$ has to satisfy the consistency condition
$$
S_0(x_j,x_k)=S_0(1/x_j,1/x_k)\,.
$$
By using this condition one can show that the Bethe equations are
invariant under the particle-to-antiparticle transformation
accompanied by changing the sign of the charge $J$.

Equation (\ref{S0}) rewritten for the dressing factor (\ref{dr})
takes the following form \bea\la{se}
\s(x_j,x_k)\s(1/x_j,x_k)=h(x_j,x_k)^{2}\,, \eea where the function
$h$ is \bea\la{Janh} h(x_j,x_k) = {x^-_k\ov x^+_k}{( 1 - {1\ov
x^-_j x^-_k} )(x^-_j - x^+_k)\ov (1 - {1\ov x^+_j x^-_k} )
  (x^+_j - x^+_k)}\,.
\eea Here $h$ is related to $f$ as follows $$ h(x_j,x_k) =
{x^-_k\ov x^+_k}f(x_j,x_k)^{-1}\,. $$ Equation (\ref{se}) admits
different solutions which should correspond to string S-matrices
in different gauges preserving the ${\rm SU}(2|2)\times {\rm
SU}(2|2)$ symmetry. This can be seen, for instance, by comparing
the string S-matrices in the temporal \cite{AF} and the light-cone
gauges \cite{AFlc,FPZ}. The light-cone Bethe equations \cite{FPZ}
have the same form as eq.(\ref{Bethe}) with
$L=P_++\sfrac{\ss+1}{2}M$, where the light-cone momentum $P_+$ is
defined as $P_+=(E+J)/2$. One can see that the temporal and
light-cone gauge string S-matrices differ by dressing factors
only; the ratio of the dressing factors satisfies eq.(\ref{se})
with $h=1$.

In spite of an attractive picture of the Hopf algebra symmetry
leading to the tight constraints on the string S-matrix at present
we do not have  any firm evidence that this is indeed the case.
Thus, we would like to confront equation (\ref{se}) against the
known leading terms in the asymptotic (strong coupling) expansion
of the dressing factor.

\medskip

The strong coupling expansion in the parametrization chosen is
simply an expansion in powers of $\eps=\frac{2\pi}{\sqrt{\l}}$
with the variable $x$ kept fixed. It is more convenient to come to
the logarithmic version of eq.(\ref{se}) which reads as \bea
i\theta(x_j,x_k)+i\theta(1/x_j,x_k)=2\log h(x_j,x_k)\, .
\label{log}\eea Then expanding the function $\log h$, we get
\bea\la{Jfexp} 2\log h(x_j,x_k) &=& -\eps {4 i x_k (x_k + x_j (-2
+ x_j x_k))\ov (x_j - x_k)(x_j x_k-1)(x_k^2-1 )}
\\\nonumber &&~~~~+\eps^2 {4 x_j^2 x_k^2(1 - 4 x_j x_k + x_j^2 +
x_k^2 + x_j^2x_k^2) \ov (x_j^2-1)(x_k^2-1 )(x_j -
x_k)^2(x_jx_k-1)^2 } + \cdots\,. \eea In order to make a
comparison of this expansion with the one of the l.h.s. of
eq.(\ref{log}) we have first to perform the sums in eq.(\ref{dph})
defining the dressing phase. Substituting in eq.(\ref{dph}) the
explicit form (\ref{charge}) of the charges we see that the
dressing phase acquires the following form \bea\la{dress2}
\theta(x_j,x_k) &=& {1\ov
\eps}\Big[\chi(x_j^-,x_k^-)-\chi(x_j^-,x_k^+)-\chi(x_j^+,x_k^-)
+\chi(x_j^+,x_k^+)\\\nonumber &&~~~~~
-\chi(x_k^-,x_j^-)+\chi(x_k^+,x_j^-)+\chi(x_k^-,x_j^+)-\chi(x_k^+,x_j^+)
\Big]\, ,
 \eea
where we have introduced \bea \chi(x,y) = -\sum_{r=2}^\infty
\sum_{n=0}^\infty {c_{r,r+1+2n}(\eps)\ov (r-1)(r+2n)}{1\ov x^{r-1}
y^{r+2n}}=\chi_0+\eps \chi_1+\cdots\, . \eea Using the explicit
form of the coefficients $c_{r,s}$ we get for the leading term
\bea\la{afs} \chi_0(x,y) = -{1\ov y} -{xy -1\ov y} \log\left({xy
-1\ov xy}\right) \, . \eea Using this formula we develop the
expansion of the l.h.s. of (\ref{log}) up to the second order in
$\eps$: \bea\la{theta0} i\theta_0(x_j,x_k) +
i\theta_0(1/x_j,x_k)=-\eps {4 i x_k (x_k + x_j (-2 + x_j x_k))\ov
(x_j - x_k)(x_j x_k-1)(x_k^2-1 )}+{\cal O}(\zeta^3)\, . \eea The
expression above literally  coincides with the leading term on the
r.h.s. of eq. (\ref{log}). Note that the subleading term of order
$\eps^2$ is absent in this expansion!

\medskip

Further, performing the sums in the first subleading correction we
get \bea\la{hl} \chi_1(x,y) &=& {1\ov \pi}\Big[\,
\log\frac{y-1}{y+1} \log\frac{x-\frac{1}{y}}{x-y}\\\nonumber
&+&{\rm
Li}_2\frac{\sqrt{y}-\sqrt{\frac{1}{y}}}{\sqrt{y}-\sqrt{x}}-{\rm
Li}_2\frac{\sqrt{\frac{1}{y}}+\sqrt{y}}{\sqrt{y}-\sqrt{x}}+{\rm
Li}_2\frac{\sqrt{y}-\sqrt{\frac{1}{y}}}{\sqrt{y}+\sqrt{x}}-{\rm
Li}_2\frac{\sqrt{y}+\sqrt{\frac{1}{y}}}{\sqrt{y}+\sqrt{x}}\, \Big]
\,. \eea This formula was obtained under the assumption that
$|xy|>1$ and \mbox{${\rm Re}(\sqrt{x}\sqrt{y})>1$.} It is then
analytically continued to the complex planes of $x$ and $y$
variables. As the function of two complex variables it has a
rather complicated structure of singularities, in particular a
branch cut singularity at $y=1$.

Again substituting this function into the dressing phase and
expanding it up to the second order in $\eps$ we find
\bea\la{theta1} &&i\theta_1(x_j,x_k) +
i\theta_1(1/x_j,x_k)\\\nonumber &&~~~~~~~~~={4i}{\pa^2\ov \pa x_j
\pa x_k}\left( \chi_1(x_j,x_k)- \chi_1(x_k,x_j)+ \chi_1(1/x_j,x_k)
- \chi_1(1/x_k,x_j)     \right) \delta x_j \delta x_k\,, \eea
where
$$
\delta x= \eps {i x^2\ov x^2-1}\,.
$$
Performing the differentiation and combining the logarithmic terms
we obtain the following result \bea \nonumber i\theta_1(x_j,x_k) +
i\theta_1(1/x_j,x_k)=\frac{i}{\pi}W(x_j,x_k)\delta x_j\delta x_k
\, ,\eea where \bea\nonumber
W(x_j,x_k)&=&4\frac{(1-4x_jx_k+x_k^2+x_j^2+x_j^2x_k^2)}
{(x_j-x_k)^2(1-x_jx_k)^2}\Big(\log\frac{x_j-1}{x_j+1}-\log\frac{1-x_j}{1+x_j}\Big)\\
\nonumber
&-&\frac{2(1+x_jx_k)}{\sqrt{x_jx_k}(1-x_jx_k)^2}\log\frac{-1+\sqrt{\frac{x_k}{x_j}}}{1+\sqrt{\frac{x_k}{x_j}}}
+\frac{1-\sqrt{x_jx_k}}{\sqrt{x_jx_k}(1-\sqrt{x_jx_k})^3}\log\frac{-x_j+\sqrt{x_jx_k}}{x_j+\sqrt{x_jx_k}}\\
&-&
\frac{1+\sqrt{x_jx_k}}{\sqrt{x_jx_k}(1+\sqrt{x_jx_k})^3}\log\frac{x_j+\sqrt{x_jx_k}}{-x_j+\sqrt{x_jx_k}}\nonumber \\
&-&\Big(\frac{2(x_j+x_k)}{(x_j-x_k)^2\sqrt{x_jx_k}}+\frac{x_j(-x_k+\sqrt{x_jx_k})}{x_k(-x_j+\sqrt{x_jx_k})^3}\Big)
\log\frac{1+\sqrt{x_jx_k}}{-1+\sqrt{x_jx_k}}\nonumber\\
&-&\frac{x_j(x_k+\sqrt{x_jx_k})}{x_k(x_j+\sqrt{x_jx_k})^3}\log\frac{-1+\sqrt{x_jx_k}}{1+\sqrt{x_jx_k}}\,
. \eea Note a non-trivial cancellation of all the terms which do
not involve logarithms. The expression for $W(x_j,x_k)$ can be
further simplified to produce the following result: \bea\nonumber
W(x_j,x_k)&=&4\frac{(1-4x_jx_k+x_k^2+x_j^2+x_j^2x_k^2)}
{(x_j-x_k)^2(1-x_jx_k)^2}\Big(\log\frac{x_j-1}{x_j+1}-\log\frac{1-x_j}{1+x_j}\Big)\\
&=&4\pi i\frac{(1-4x_jx_k+x_k^2+x_j^2+x_j^2x_k^2)}
{(x_j-x_k)^2(1-x_jx_k)^2} \, , \eea where we used the principle
branch of $\log$. Thus, we finally arrive at \bea
i\theta_1(x_j,x_k) + i\theta_1(1/x_j,x_k)=-
4\frac{(1-4x_jx_k+x_k^2+x_j^2+x_j^2x_k^2)}
{(x_j-x_k)^2(1-x_jx_k)^2} \delta x_i \delta x_j\, .\eea One can
now recognize that this expression perfectly matches the $\eps^2$
term in the r.h.s. of eq.(\ref{log}). It is interesting to note
that if we would drop all ${\rm Li}_2$-functions in eq.(\ref{hl})
keeping only the logarithms we would still satisfy eq.(\ref{log})
at order $\eps^2$. However, dilogarithmic functions are necessary
for the dressing phase to be expandable in Taylor series in local
excitation charges. This is clearly related to yet to be
understood analytic properties of the dressing phase.

To summarize, we have found that the perturbative string S-matrix
satisfies the equation (\ref{se}) on the dressing factor arising
upon requiring the existence of the Hopf algebra structure up to
two leading orders in the strong coupling expansion.

There are many open interesting questions. First of all it is
unclear what additional (analyticity) conditions one should impose
to restrict the space of solutions of the functional equation
\cite{J}. Second, one would like to understand how the
representation used in \cite{B} to derive the S-matrix from the
symmetry algebra might appear by quantizing string theory in a
particular gauge. In particular, one should be able to recover the
central charges introduced in \cite{B} in the symmetry algebra of
gauge-fixed string theory. The derivation of \cite{J} was based on
the existence of a Hopf algebra structure of gauge-fixed string
theory. It is important to find an origin of the structure in
string theory. Finally, it would be interesting to establish a
connection of the approach used in \cite{B,J} to that of
\cite{MP}.

\section*{Acknowledgments }

We are  grateful to R. Janik, J. Plefka, M. Staudacher, A.
Tseytlin and M. Zamaklar for useful
 discussions and comments.
  The work of
G.~A. was supported in part by the European Commission RTN
programme HPRN-CT-2000-00131, by RFBI grant N05-01-00758, by NWO
grant 047017015 and by the INTAS contract 03-51-6346. The work of
S.~F.~was supported in part by the EU-RTN network {\it
Constituents, Fundamental Forces and Symmetries of the Universe}
(MRTN-CT-2004-005104).


\end{document}